\documentclass{iopconfser}

\usepackage[]{xcolor}
\usepackage[most]{tcolorbox}
\usepackage{amsmath}

\usepackage[utf8]{inputenc}
\usepackage{csquotes}
\usepackage[backend=bibtex,sorting=none, doi=false,isbn=false,url=false,eprint=false]{biblatex}
\usepackage{amssymb}
\usepackage{txfonts}
\usepackage[english]{babel}
\usepackage{mathtools}
\usepackage{physics}
\usepackage{multicol}
\usepackage{graphicx}
\usepackage{wrapfig}
\usepackage[margin=5pt,font=small,labelfont=bf,labelsep=endash]{caption}
\usepackage{placeins}
\usepackage{slashed}
\usepackage{lipsum}
\usepackage{feynmf}
\usepackage{array}
\usepackage[thinlines]{easytable}
\usepackage{makecell}
\usepackage{hyperref}
\usepackage{enumitem}
\usepackage{pifont}

\DeclareFieldFormat[article]{volume}{\textbf{#1}}
\renewbibmacro{in:}{}

\DeclareBibliographyDriver{article}{%
  \printnames{author} 
\space 
  \printfield{year} 
\space 
  \printfield{title} 
\space 
  \printfield{journaltitle} 
\space 
  \printfield{volume} 
\printfield{number}
  \printfield{pages} 
  \printfield{addendum} 
}

\begin{document}

\definecolor{mypurple}{cmyk}{0.5, 0.7808, 0, 0.1412}
\definecolor{mypink}{cmyk}{0, 0.7808, 0.4429, 0.1412}
\definecolor{myblue}{cmyk}{1, 0.5, 0, 0.1412}
\definecolor{myindigo}{cmyk}{0.8, 1, 0, 0}
\definecolor{mycyan}{cmyk}{1, 0, 0.2, 0.1}
\definecolor{myorange}{cmyk}{0, 0.333, 1, 0}
\definecolor{myorange2}{cmyk}{0.3333, 0.666, 1, 0}

\title{Bi-Hamiltonian Structures and Equivalent Representations of the Pais-Uhlenbeck Model.}

\author{Bethan Turner$^1$}

\affil{$^1$Department of Mathematics, City St George's, University of London, London, UK}
\email{bethan.turner.2@city.ac.uk}

\begin{abstract}

We provide a complete classification of all the ways the Pais-Uhlenbeck osicllator might be embedded in two dimensional space. We discuss the Bi-Hamiltonian structures of this model, and examine how alternative Hamiltonian structures might be generated from the dynamical Lie symmetries of the theory. We then examine how the Bi-Hamiltonian strucutre may be exploited to evade the problem of unbounded Hamiltonians that is usually associated with Higher Time Derivative Theories. The effect of interactions on this Bi-Hamiltonian structure is also considered.  
\end{abstract}

\section{Introduction}
\label{intro}

The question of whether or not a viable physical theory can be a Higher Time Derivative Theory (HTDT), meaning a theory where the Lagrangian is at least second order in time derivatives, continues to be of some interest. While such theories are generally discounted as unphysical on the grounds that they contain fundamental instabilities associated with the absence of a lower bound on the Hamiltonian, they continue to be of interest due to the fact that they tend to be renormalisable, most notably in the case of higher derivative quantum gravity \cite{stelle}. Other potential benefits are the avoidance of the self interaction problem in classical electromagnetism \cite{Podolsky,Gratus_2015}, or the avoidance of the cosmological singularity \cite{Starobinsky1980}. 

There is a price to pay for these appealing properties however. In \cite{stelle}, it is  shown that the renormalisable gravity theory comes at the cost of the introduction of either negative norm or negative energy states, a charecteristic feature of quantum HTDTs.  The model proposed in \cite{Starobinsky1980} is known to lead to the unstable perturbing of flat space \cite{simon92}. The model proposed in  \cite{Podolsky} leads to an energy momentum tensor with a negative part \cite{Gratus_2015} . The fundamental reason for these problems is that, even at the classical level, the energy does not have a global minimum  \cite{Orstogradskytheorem}. This in turn leads to a quantum picture where an unbounded energy eigenvalue spectrum can only be avoided at the price of the introduction of negative norm states, indeed this is what happens in \cite{stelle}. 

Given the appealing properties of HTDTs, a large amount of work has been done on solving the problems discussed above. Ideas include imposing constraints on the system \cite{Exorcism},  restricting our interest to low energy effective theories \cite{Hawking_2002} and distinguishing between benign and malicous theories, where we search for systems with stable dynamics despite the presence of higher time derivatives \cite{Smilga_2017,Smilga_2005}. 

One of the most prototypical HTDTs is a system known as the Pais-Uhlenbeck oscillator, first studied in \cite{paisuhlenbeck}. There is already a large amount written about this model, including many works on how the "ghost problem" (the name typically given to the problems associated with HTDTs) might be resolved and/or circumvented. For instance it is sometimes suggested that it is possible to embed the model in a two dimensional space in such a way as to preserve the dynamics but give a positive definite Hamiltonian \cite{mostafazadeh,Nucci_2010,stephen}. Other suggestions include generating a positive definite Hamiltonian from the conserved quantities of the model \cite{Nucci2011}, or that the existence of a positive definite conserved quantity stabilises the model \cite{Kaparulin2014}. Many other works have of course been written, but it is not possible to cite them all here. 

In this article we will review and classify all the different ways this higher derivative model may be equivalently represented as a first order two dimensional model. We will also show how the Bi-Hamiltonian structure of this model, which as far as we can tell was first derived in \cite{Damaskinsky_2006}, can be found via consideration of the Lie symmetries of this model. I will then discuss how attempts to write down a positive definite two dimensional Hamiltonian of this system only make sense in relation to the Bi-Hamiltonian structure. In the last section I will review how this is changed by the inclusion of interactions. A more detailed discussion of the work in this paper can be found in \cite{felski2025}. 

\section{The Pais-Uhlenbeck Oscillator, Classical Dynamics and Quantisation}

The Pais-Uhlenbeck oscillator is probably one of the most extensively studied HTDTs, beginning with its first consideration in \cite{paisuhlenbeck}. In its simplest version it is a linear theory that is quadratic in derivatives such that the lagrangian and equation of motion are given by \\
\begin{subequations}
\begin{minipage}{0.49\linewidth}
\begin{align}
L  = \frac{1}{2} \left( \beta q^2 - \alpha \dot{q}^2 + \ddot{q}^2 \right)
\label{PUlagrangian}
\end{align}
\end{minipage}
\begin{minipage}{0.49\linewidth}
\begin{align}
0  = q^{(4t)} +  \alpha \ddot{q} + \beta q,
\label{PUeom}
\end{align}
\end{minipage}
\label{PUconfig}
\end{subequations} \\
where we define the parameters $\alpha  :=  \omega_1^2 + \omega_2^2$,  $\beta := \omega_1^2 \omega_2^2$,  $\omega_1, \omega_2 \epsilon \mathbb{R}$.

The phase space of this system can be constructed using an old method devised by Ostorgradsky in the $19^{th}$ century, which gives a higher derivative generalisation of the Legendre transform, see \cite{Ostrogradsky1850,Orstogradskytheorem}. From the four dimensional configuration space described in (\ref{PUconfig}), we get a four dimensional phase space with coordinates
\begin{subequations}
\begin{minipage}{0.2\linewidth}
\begin{align}
X_1 = & q  
\label{X1}
\end{align}
\end{minipage}
\begin{minipage}{0.2\linewidth}
\begin{align}
X_2 = & \dot{q}
\label{X2}
\end{align}
\end{minipage}
\begin{minipage}{0.35\linewidth}
\begin{align}
P_1 & =  \frac{\delta L}{\delta \dot{q}} =  - \alpha \dot{q} - q^{(3t)} 
\label{P1}
\end{align}
\end{minipage}
\begin{minipage}{0.25\linewidth}
\begin{align}
P_2 & = \frac{\delta L}{\delta \ddot{q}} =  \ddot{q}.
\label{P2}
\end{align}
\end{minipage}
\label{PUphase}
\end{subequations}

From the coordinates in (\ref{PUphase}) we can construct the Hamiltonian and Poisson tensor 
\begin{align}
J_1   =  \frac{\partial }{\partial X_1} \wedge \frac{\partial }{\partial P_1} + \frac{\partial }{\partial X_2} \wedge \frac{\partial }{\partial P_2}, = - \frac{\partial}{\partial q} \wedge \frac{\partial}{\partial q^{(3)}} + \frac{\partial}{\partial \dot{q}} \wedge \frac{\partial}{\partial \ddot{q}} +  \alpha \frac{\partial}{\partial \ddot{q}}  \wedge \frac{\partial}{\partial q^{(3)}}  
\label{PUpoissontensor}
\end{align}
\begin{align}
H_1  = P_2 \ddot{q}\left( P_1, P_2, X_1, X_2 \right) + P_1 X_2 - L = - \dot{q} q^{(3)}  + \frac{1}{2} \ddot{q}^2 - \frac{1}{2} \beta q^2  -  \frac{1}{2} \alpha \dot{q}^2.
\label{PUhamiltonian}
\end{align}
From the Hamiltonian we can see the origin of the problem with HTDTs. It is linear in the coordinate $P_1$, meaning that, as mentioned above, it does not posses a global minimum. This is what is thought to be the origin of the instability, allowing for runaway solutions even at the classical level. What is noteworthy about the Pais-Uhlenbeck oscillator is that in its free form it does not exhibit runaway behaviour. In fact, as we will now see, if we solveequation (\ref{PUeom}) we find oscillatory solutions that essentially matches the sum of two harmonic oscillators, 
\begin{align}
q =  a_{1}^{*} e^{i \omega_1 t} + a_{1} e^{- i \omega_1 t} + a_{2} e^{- i \omega_2 t} + a_{2}^{*} e^{i \omega_2 t} &,&
 a_i =  a_i \left(q, \dot{q}, \ddot{q}, q^{(3t)} \right) \big \rvert_{t=0}.
 \label{PUsolution}
\end{align}
Therefore in its free form this theory is perfectly well behaved. This can change in the presence of interactions, but this issue will be dealt in section \ref{interactions}. 

Before proceeding a brief comment on the quantum theory is warranted. We canonically quantise in the usual way, promoting the coordinates to operators $q^{i} \lvert_{t=0} \to \hat{q}^{(i)}$,  which in turn means the coefficients in (\ref{PUsolution}) are also promoted to operators,  $a_i \to \hat{a}_i$, $a^*_i  \to  \hat{a}_i^\dagger $.   The classical Poisson brackets are also promoted to commutators, allowing us to derive the relations, \\
\begin{subequations}
\begin{minipage}{0.19\linewidth}
\begin{align}
\left[ \hat{q}^{(1)}, \hat{q}^{(2)} \right] =   i,
\end{align}
\end{minipage}
\begin{minipage}{0.19\linewidth}
\begin{align}
 \left[ \hat{q}, \hat{q}^{(3)} \right] = - i,
\end{align}
\end{minipage}
\begin{minipage}{0.2\linewidth}
\begin{align}
 \left[ \hat{q}^{(2)}, \hat{q}^{(3)} \right]  = i  \alpha,
\end{align}
\end{minipage}
\begin{minipage}{0.2\linewidth}
\begin{align}
 \left[ \hat{a}_{1} , \hat{a}_{1}^{\dagger} \right]  =   \frac{1}{2 R_1}
\end{align}
\end{minipage}
\begin{minipage}{0.22\linewidth}
\begin{align}
\left[ \hat{a}_{2} , \hat{a}_{2}^{\dagger} \right] =  - \frac{1}{2 R_2},
\end{align}
\end{minipage}
\end{subequations}
where $R_{i}  := \omega_i^2  \left( \omega_1^2 -\omega_2^2 \right)$. The quantum Hamiltonian can then be expressed as 
\begin{align}
\hat{H}= 2  \left( R_1 \hat{a}_{1}^{\dagger} \hat{a}_{1} - R_2 \hat{a}_{2}^{\dagger} \hat{a}_{2} \right) + \frac{1}{2} \left( \omega_1 +\omega_2 \right). 
\label{PUquantumHamiltonian}
\end{align}

To see the quantum analogue of the classical problem discussed above, our two options for the choice of ground state. In the first option, we treat both modes $\omega_1$ and $\omega_2$ as positive energy modes, such that the ground state vanishes under the action of both annihalation operators
\begin{align}
 \hat{a}_{i} \ket{0} & = 0, & \bra{0}  \hat{a}_{i} \hat{a}_{i}^{\dagger} \ket{0} & = \frac{(-1)^{i-1}}{2 R_i}, &  \hat{H} \hat{a}_{i}^{\dagger} \ket{0} & = \frac{\left( \omega_1 + \omega_2 \right)}{2} + \omega_i.
 \end{align}
 In this choice, the Hamiltonian eigenvalue spectrum is bounded, the ground state energy is the lowest energy. However, it also means we have negative norm states, such as the one shown in for (\ref{quantneg}). Alternatively, we can choose $\omega_1$ and $\omega_2$ to be positive and negative energy modes respectively, in which case we have,
\begin{subequations}
\begin{align}
 \hat{a}_{1} \ket{0} & = 0, & \bra{0}  \hat{a}_{1} \hat{a}_1^{\dagger} \ket{0} & = \frac{1}{2 R_1}, &  \hat{H} \hat{a}_2 \ket{0} & = \frac{ \left( 3 \omega_1 - \omega_2 \right) }{2} \\
   \hat{a}_2^{\dagger} \ket{0} &= 0, & \bra{0}  \hat{a}_2^{\dagger} \hat{a}_2 \ket{0} &=   \frac{1}{2 R_2}, &  \hat{H} \hat{a}_2 \ket{0} & = \frac{\left(  \omega_1 - 3 \omega_2 \right) }{2}, 
\label{quantneg}
 \end{align}
\end{subequations}
 so we avoid the negative norm states but the eigenvalue spectrum is unbounded. There are energies lower than that of the "ground state", so it is not even really accurate to call it a ground state anymore. Neither of these scenario's are particularly desirable. 

\section{Two Dimensional Representations}
\label{Twodimensionalreps}

It is often suggested that it is possible to solve this ghost problem by writing some two dimensional theory that has the same dynamics as the Pais-Uhlenbeck oscillator. If this theory has a positive definite Hamiltonian, the argument goes, the problem is therefore solved, as this positive definite Hamiltonian can be quantised instead. Versions of this argument are put forward in, for example, the papers by \cite{mostafazadeh, stephen,Nucci_2010}. In this section we consider the most generic possible version of this argument.  We start with a general two dimensional first order Lagrangian and two corresponding second order Euler-Lagrange equations \\
\begin{subequations}
\begin{minipage}{0.44\linewidth}
\begin{align}
L_{fo} = \frac{a_x}{2} \dot{x}^2 + \frac{a_y}{2} \dot{y}^2 - \frac{b_x}{2} x^2 - \frac{b_y}{2} y^2 - g x y 
\label{Lagrangiantwodimensions}
\end{align}
\end{minipage}
\begin{minipage}{0.28\linewidth}
\begin{align}
  a_x \ddot{x} + b_x x + g y = 0, 
\label{twodimeom1}
\end{align}
\end{minipage}
\begin{minipage}{0.28\linewidth}
\begin{align}
  a_y \ddot{y} + b_y y + g x = 0.
\label{twodimeom2}
\end{align}
\end{minipage}
\label{twodimensionsconfig}
\end{subequations}
We want to map the dynamics of the system in (\ref{twodimensionsconfig}) to those of the Pais-Uhlenbeck oscillator. There are two options for how to do this. Consider some map from the coordinates $x$ and $y$ to a some linear combination of $q$ and $\ddot{q}$
\begin{align}
 f_{i}:  \left(x, y \right) \to \left( q, \ddot{q} \right)&,&    i = a,b.
\end{align}
The index $a$ and $b$ refer to our two choices of maps. We want to map the dynamics of the equations in (\ref{twodimensionsconfig}) to the dynamics in (\ref{PUconfig}). This means that we have to map either one or both of the equations of motion in  (\ref{twodimensionsconfig}) to the one in (\ref{PUconfig}), as summarised below
\begin{subequations}
\begin{align}
f_a & : \phi_1  \to \phi_{PU}, &  f_a & : \phi_2  \to \phi_{PU}, \label{mapsa} \\
f_b & : \phi_1   \to \phi_{PU}, & f_b  &: \phi_2  \to 0. \label{mapsb}
\end{align}
\end{subequations}
Note that $\phi_{PU}$ represents the Pais-Uhlenbeck equation as given in equation (\ref{PUeom}), while $\phi_1$ and $\phi_2$ are the equations (\ref{twodimeom1}) and \ref{twodimeom2} respectively. We find these maps by writing down some generic coordinate transform, 
\begin{align}
x  = \mu_0 q + \mu_2 \ddot{q} &,& y  = \nu_0 q + \nu_2 \ddot{q}, \label{coordinates}
\end{align}
substituting these into (\ref{twodimensionsconfig}) and solving the resulting equations. From both maps $a$ and $b$ we get two sets of solutions, as shown below. 

\begin{align}
\mathbf{Ta_1^{\pm}} &\mathbf{:} & a_x & \neq 0, & b_x^{\pm} & = \frac{a_x}{2} \left( \alpha - \frac{2 g}{a_y} + \rho^{\pm}_0 \right), & \mu_0^{\pm}  & = \frac{1}{2 a_x} \left( \alpha + \rho_0^{\pm} \right), & \mu_2^{\pm}  & = \frac{1}{2 a_x} \nonumber \\
& & a_y & \neq 0, & b_y^{\pm} & = \frac{a_y}{2} \left( \alpha - \frac{2 g}{a_x} + \rho^{\pm}_0 \right),  & \nu_0^{\pm}  & = \frac{1}{2 a_y} \left( \alpha + \rho_0^{\pm} \right), & \nu_2^{\pm}  & = \frac{1}{2 a_y}, \nonumber  \\
\nonumber  \\
\mathbf{Ta_2^{\pm}} &\mathbf{:} & a_x & \neq 0, & b_x^{\pm}  & = \frac{ \left( \alpha + \rho_g^{\pm} \right)}{2 a_x}, & \mu_0^{\pm} & = \frac{1}{2 a_x} \left( \alpha - \frac{2 g}{a_y} + \rho^{\pm}_g \right), & \mu_2^{\pm}  & = \frac{1}{2 a_x}, \nonumber \\
& & a_y & \neq 0, & b_y^{\pm}  & = \frac{ \left( \alpha + \rho_g^{\pm} \right)}{2 a_y}, & \nu_0^{\pm} & = \frac{1}{2 a_y} \left( \alpha - \frac{2 g}{a_x} + \rho^{\pm}_g \right), & \nu_2^{\pm} & = \frac{1}{2 a_y},   \nonumber  
\end{align}

\begin{align}
\mathbf{Tb_1^{\pm}} &\mathbf{:} & a_x & \neq 0, & b_x  & \neq \frac{a_x}{2} \left( \alpha + \rho_0^{\pm} \right), & \mu_0^{\pm} & = \frac{1}{ a_x} \left( \alpha - \frac{b_x}{a_x}  \right), & \mu_2^{\pm}  & = \frac{1}{a_x}, \nonumber \\
& &a_y & = - \frac{a_x g}{\tau^2}, & b_y  & = \frac{g}{\tau} \left( b_x - a_x \alpha \right), & \nu_0^{\pm} & = \frac{\tau}{g a_x^2}, &  \nu_2^{\pm}  & = 0, \nonumber \\
\nonumber  \\
\mathbf{Tb_2^{\pm}} &\mathbf{:} & a_x & \neq 0, & b_x^{\pm}  & = \frac{g^2}{b_y} +  \frac{a_x}{2} \left( \alpha + \rho_0^{\pm} \right), & \mu_0^{\pm} & = \frac{2 \beta }{ a_x \left( \alpha + \rho^{\pm} \right)}, & \mu_2^{\pm}  & = \frac{1}{a_x} \nonumber \\
& &a_y & = 0, & b_y^{\pm}  & \neq 0, & \nu_0^{\pm} & = \frac{2 \beta g}{ a_x b_y \left( \alpha + \rho^{\pm}_0 \right)}, & \nu_2^{\pm}  & = - \frac{g}{ a_x b_y}   \nonumber 
\end{align}
where for the sake of neatness we have defined the constants
\begin{align}
\rho_g^{\pm} : =  \pm  \sqrt{ \alpha^2 - 4 \beta - \frac{4 g^2}{a_x a_y}} &, & \rho_0^{\pm} = \rho_{g=0}^{\pm}&, &  \tau := b_x^2 - a_x b_x \alpha + a_x^2 \beta \nonumber. 
\end{align}

Two of these solutions $\mathbf{Ta_1^{\pm}}$ and $\mathbf{Tb_2^{\pm}}$ are not of particular interest. In $\mathbf{Ta_1^{\pm}}$ the solution is such that the coordinates $x$ and $y$ are the same up to multiplication by an overall constant, thus the transform is singular. In $\mathbf{Ta_2^{\pm}}$ $a_y$ is vanishing, which would make the Lagrangian in (\ref{Lagrangiantwodimensions}) degenerate. Therefore in the following argument we mainly focus on $\mathbf{Ta_2^{\pm}}$ and $\mathbf{Tb_1^{\pm}}$. From these two solutions we find the following two Hamiltonians,
\begin{align}
H_{Ta_2^{\pm}} & = \frac{a_x }{2} p_x^2 + \frac{a_y }{2} p_y^2 + \frac{\alpha + \rho_g^{\pm}}{2 a_x} x^2 +  \frac{\alpha + \rho_g^{\mp}}{2 a_y} y^2 + g x y. \\
H_{Tb_1^{\pm}} & = \frac{a_x }{2} p_x^2 + \frac{a_x g }{2 \tau^2} p_y^2 + \frac{\alpha + \rho_0^{\pm}}{2 a_x} x^2 + \frac{g \left( b_x - a_x \alpha \right)}{\tau} y^2 + g x y.
\end{align}
From this it would seem that if we simply fix the remaining free parameters so that these Hamiltonians are positive definite, we have the solution to the problem, this is certainly the argument of \cite{mostafazadeh}. However there is a problem, which we can see by consider the Poisson tensor. If  we map the two dimensional Poisson tensor to its higher derivative version, we get
\begin{align}
f_i: \frac{\partial }{\partial x} \wedge \frac{\partial }{\partial p_x} \wedge \frac{\partial }{\partial y} \wedge \frac{\partial }{\partial p_y}  \to \frac{1}{\mu_0 \nu_2 - \mu_2 \nu_0} & \left[  \left( \frac{1}{a_x} \nu_2^2 + \frac{1}{a_y} \mu_2^2 \right) \frac{\partial }{\partial q} \wedge \frac{\partial }{\partial \dot{q} }+ \left(-\frac{1}{a_x} \nu_0 \nu_2 - \frac{1}{a_y} \mu_0 \mu_2 \right) \left( \frac{\partial }{ \partial q} \wedge \frac{\partial }{\partial q^{(3)}} - \frac{\partial }{\partial \dot{q} }  \wedge \frac{\partial }{\partial \ddot{q} }   \right) \right. \nonumber \\ & \hspace{0.5cm}  \left.  + \left( \frac{1}{a_x} \nu_0^2 + \frac{1}{a_y} \mu_0^2  \right) \frac{\partial }{\partial \ddot{q}} \wedge \frac{\partial }{\partial q^{(3)}} \right]. 
\label{Poissontwodimensions}
\end{align}
Note that we have not yet implemented any of the above solutions. However if we compare the Poisson tensors in (\ref{Poissontwodimensions}) and (\ref{PUpoissontensor})we see that for these to match, then the term $\frac{1}{\mu_0 \nu_2 - \mu_2 \nu_0}   \left( \frac{1}{a_x} \nu_2^2 + \frac{1}{a_y} \mu_2^2 \right) \frac{\partial }{\partial q} \wedge \frac{\partial }{\partial \dot{q} }$  in (\ref{Poissontwodimensions}) needs to vanish. This is only possible if, $a_x$ and $a_y$ are of opposite sign. However, this means we have a kinetic part with terms of opposite sign. Thus, althought it appears as a diffrence between two squares not as the result of a linea term as in (\ref{PUhamiltonian}), we have an unbounded Hamiltonian. Thus if we wish to preserve the Poisson tensor we cannot avoid the ghost problem. If we change the Poisson tensor, this amounts to essentially changing the problem, especially in the quantum case. Since canonical quantisation rests on promoting the Poisson brackets to commutators, if we change the Poisson tensor then we will in effect be quantising another model. The only way to circumvent this problem is to search for some alternative Hamiltonian structure. 

\section{Lie Symmetires and Bi-Hamiltonian Structures}

In this section we review the Bi-Hamiltonian structure of the Pais-Uhlenbeck oscillator and discuss how it may be generated from the dynamical Lie symmetries. We start by writing down the vector field
\begin{align}
V  = J_1 \left( . , dH_1 \right) = \dot{q} \partial_q + \ddot{q} \partial_{\dot{q}} + q^{(3)} \partial_{\ddot{q}} - \left( \alpha \ddot{q} + \beta q \right) \partial_{q^{(3)}}. 
\label{vectorfield}
\end{align}
We generate the Lie symmetries of the flow this theory by means of a point transform, see for instance \cite{Hydon_2000} where we vary the position coordinates with respect to some infintesimal parameter $\varepsilon$ so that 
\begin{align}
q^{(i)} \to & \tilde{q}^{(i)} + \varepsilon \xi_i
\end{align}
which in turns leads to the following variation in the vector field \\
\begin{subequations}
\begin{minipage}{0.5\linewidth}
\begin{align}
\frac{d \tilde{q}^{(i)} }{dt} =  V_i \left( \tilde{q}^{(i)} \right) - \delta V_i  \left( \tilde{q}^{(i)} \right),
\end{align}
\end{minipage}
\begin{minipage}{0.5\linewidth}
\begin{align}
\delta V_i  \left( \tilde{q}^{(i)} \right) =  \varepsilon \left[ X, V \right]_i \left(\tilde{q}^{(i)} \right) = 0.
\label{varV}
\end{align}
\end{minipage}
\end{subequations}
The vector fields $X$ are the Lie symmetries of the flow. Assuming these are linear, we can make the generic anstaz
\begin{align}
X  = & \sum_{i=1}^4 \xi_{i} \partial_{q^{(i)}} \nonumber
\end{align}
and substitute this into (\ref{varV}). Solving the resulting equations for $\xi_i$ we find the four Lie symmetries
\begin{align}
X_1  = &  \dot{q} \partial_q + \ddot{q} \partial_{\dot{q}} + q^{(3)} \partial_{\ddot{q}} - \left( \alpha \ddot{q} + \beta q \right)  \partial_{q^{(3)}}, &
X_2 = & \frac{1}{2} \left[ q \partial_q + \dot{q} \partial_{\dot{q}} + \ddot{q} \partial_{\ddot{q}} + q^{(3)} \partial_{q^{(3)}} \right],\nonumber \\
X_3  = & \frac{1}{2} \left[ \ddot{q} \partial_q + q^{(3)} \partial_{\dot{q}} - \left( \alpha \ddot{q} + \beta q \right) \partial_{\ddot{q}} - \left( \alpha q^{(3)} + \beta \dot{q} \right)  \partial_{q^{(3)}} \right], &
X_4  = &  \left( \alpha q + q^{(3)} \right) \partial_q - \beta \left( q \partial_{\dot{q}} + \dot{q} \partial_{\ddot{q}} + \ddot{q}  \partial_{q^{(3)}} \right). \nonumber 
\end{align} 
$X_1$ is the vector field, $X_2$ is the Euler operator. It can be straightforwardly verified that \begin{align}
\left[ X_i, X_j \right]  & = 0 \hspace{0.5cm} i,j=1,2,3,4, \nonumber
\end{align}
thus the Lie algebra formed by $X_i$ is Abelian. Their actions on the Hamiltonian $H_1$ are shown below,
\begin{align}
X_1 \left( H_1 \right)  =  0 &,&  X_2 \left( H_1 \right) =  H_1 &,& X_3 \left( H_1 \right)  =   \frac{1}{2} \beta \dot{q}^2 - \frac{1}{2} \alpha \ddot{q}^2  - \frac{1}{2} \left( q^{(3)} \right)^2 - \beta q \ddot{q} &,&   X_4 \left( H_1 \right)  = 0. \nonumber
\end{align}
Crucially, under the action of $X_3$, $H_1$ is mapped to a new independent charge. If we treat this as a new Hamiltonian, we can solve the equation 
\begin{align}
J_2 \left( . , dH_2 \right) & = J_1 \left( . , dH_1 \right),
\label{BiHamiltonian}
\end{align}
to find the tensor $J_2$ and thus define an alternative Hamiltonian structure \\
\begin{subequations}
\begin{minipage}{0.5\linewidth}
\begin{align}
H_2  := q \ddot{q}  - \frac{1}{2}\dot{q}^2 - \frac{\alpha}{2 \beta} \ddot{q}^2  + \frac{1}{2 \beta} \left( q^{(3)} \right)^2,
\end{align}
\end{minipage}
\begin{minipage}{0.5\linewidth}
\begin{align}
J_2  := \frac{\partial}{\partial q} \wedge \frac{\partial}{\partial \dot{q}} +  \omega_1^2 \omega_2^2  \frac{\partial}{\partial \ddot{q}}  \wedge \frac{\partial}{\partial q^{(3)}}.
\end{align}
\end{minipage}
\end{subequations}\\
Thus this system is Bi-Hamiltonian, a concept first introduced in 
\cite{biHamiltonian1, Gelfand1979} and well known in the study of integrable systems, see for instance \cite{DAS}, although the two are not exactly the same. The existence of this Bi-Hamiltonian structure allows us to form linear combinations of the Hamiltonian and Poisson tensor, \\
\begin{subequations}
\begin{minipage}{0.5\linewidth}
\begin{align}
H'  = c_1 H_1 + c_2 H_2,
\label{Hamiltoniangeneral}
\end{align}
\end{minipage}
\begin{minipage}{0.5\linewidth}
\begin{align}
J'  = \frac{ c_1 J_1 + \omega_1^2 \omega_2^2 c_2 J_2 }{ \left( c_1 - c_2 \omega_1^2 \right)  \left( c_1 - c_2 \omega_2^2\right), } 
\label{Poissontensorgeneral}
\end{align}
\end{minipage}
\end{subequations}
In \cite{Damaskinsky_2006} a version of the tensor $J'$ was found by solving the equations resulting from taking the Lie derivative of a contravariant anti-symmetric tensor with respect to the vector field in (\ref{vectorfield})
\begin{align}
l_V \left( J \right)  = 0.
\label{liederivative}
\end{align}
 Equation (\ref{BiHamiltonian}) can then be solved to find $H_2$. This approach gives the same results as what we found in \cite{felski2025} by utillising the Lie symmetries of the flow.  Making use of these symmetries gives us a more systematic way to generate alternative Hamiltonians. The value of this approach becomes clearer if we wish to consider the higher order versions of the Pais-Uhlenbeck oscillator, as the ansatz for a generic unknown Hamiltonian becomes more unwieldy, see \cite{felski20252}. 

We now return to the question of wether or not we can find a positive definite PU Hamiltonian. To begin, we rewrite (\ref{Hamiltoniangeneral}) in the following form,
\begin{align}
H' = \sum_{i,j=1, i \neq j}^2 \frac{ \omega_i^2 }{ 2 \left( c_1 \omega_i^2 - c_2 \right) \left( \omega_i^2 -\omega_j^2 \right)} \left[ \left( q^{(3)} + \omega_j^2 \dot{q}\right)^2 + \omega_i^2 \left( \ddot{q} + \omega_j^2 q \right)^2 \right].
\label{Hamiltoniansquares}
\end{align}
In \cite{bolonek2005} they use an expression of this form as an ansatz, however as we have seen it can be systematically generated by linearly combining charges generated via the Lie symmetries as discussed above, see also  \cite{felski20252} for the $N=3$ case. We can see from (\ref{Hamiltoniansquares})  that we will have a positive definite Hamiltonian if the inequality,
\begin{align}
\left( c_1 \omega_1^2 -c_2 \right) \left( \omega_1^2 -\omega_2^2 \right) > 0 \wedge \left( c_1 \omega_2^2 -c_2 \right) \left( \omega_2^2 -\omega_1^2 \right) > 0  
\end{align}
 is satisfied. 
 To make a connection with the discussion in the previous section, we briefly consider matching $H'$ to the two dimensional models discussed above. Starting from the generic two dimensional Hamiltonian and Poisson tensor \\
\begin{subequations}
\begin{minipage}{0.5\linewidth}
\begin{align}
H_{fo}  = \frac{p_x^2}{2 a_x} + \frac{p_y^2}{2 a_y} + \frac{b_x}{2} x^2 + \frac{b_y}{2} x^2 + g x y,
\label{Hamiltoniangeneral}
\end{align}
\end{minipage}
\begin{minipage}{0.5\linewidth}
\begin{align}
J_{fo}  = \frac{\partial}{\partial x} \wedge \frac{\partial}{\partial p_x} + \frac{\partial}{\partial y} \wedge \frac{\partial}{\partial p_y}, 
\label{Poissontensorgeneral}
\end{align}
\end{minipage}
\end{subequations}
we act on these with the maps constructed in the section \ref{Twodimensionalreps}. For each of these maps , $H_{fo}$ is mapped to some linear combination of $H_1$ and $H_2$, and $J_{fo}$ is mapped to some linear combination for $J_1$ and $J_2$, \\
\begin{subequations}
\begin{minipage}{0.5\linewidth}
\begin{align}
f_i : H_{fo} = c_1 H_1 + c_2 H_2,
\label{Hamiltoniangeneral}
\end{align}
\end{minipage}
\begin{minipage}{0.5\linewidth}
\begin{align}
f_i : J_{fo}  = \frac{ c_1 J_1 + \omega_1^2 \omega_2^2 c_2 J_2 }{ \left( c_1 - c_2 \omega_1^2 \right)  \left( c_1 - c_2 \omega_2^2  \right) }, 
\label{Poissontensorgeneral}
\end{align}
\end{minipage}
\end{subequations}
For each of the maps constructed in the previous section, $c_1$ and $c_2$ are expressed as some function of the remaining free parameters, as shown below. 

\begin{align}
\mathbf{Ta_1^{\pm}} &\mathbf{:} &  c_1 & =  c_2  M^{\pm} \left( \omega_1^2,  \omega_2^2 \right), & c_2 & = - \frac{a_x + a_y}{a_x a_y} \nonumber  \\
\nonumber  \\
\mathbf{Ta_2^{\pm}} &\mathbf{:} & c_1 & = \frac{1}{2 a_x a_y}\left( 4g - \rho_g^{\pm} \right) + \frac{\alpha}{2}c_2, & c_2 & = - \frac{a_x + a_y}{a_x a_y} \nonumber \\
\nonumber \\
\mathbf{Tb_1^{\pm}} &\mathbf{:} &  c_1 & = \frac{1}{a_x} \left( \frac{b_x}{a_x} - \alpha \right), &  c_2  & = - \frac{1}{a_x} \hspace{0.8cm}  \nonumber \\
\nonumber  \\
\mathbf{Tb_2^{\pm}} &\mathbf{:} & c_1 & =  - \frac{1}{a_x} M^{\pm} \left( \omega_1^2,  \omega_2^2 \right), &  c_2 &= - \frac{1}{a_x }. \nonumber 
\end{align}
where $M^{+}\left( \omega_1^2 , \omega_2^2 \right) : =   \text{max} \left( \omega_1^2 , \omega_2^2 \right), M^{-}\left( \omega_1^2 , \omega_2^2 \right) : =   \text{min} \left( \omega_1^2 , \omega_2^2 \right) $. A few comments are in order, the first is that $\mathbf{Ta_1^{\pm}}$ and $\mathbf{Tb_2^{\pm}} $ both give a singular Poisson tensor. Given what was observed above about them giving singular transforms/ degenerate systems, this is to be expected. As we saw above, these maps do not describe a well behaved transform. Solution $\mathbf{Ta_2^{\pm}}$ allows us to preserve any combination of $H_1$ and $H_2$, depending on how we fix $a_x$ and $a_y$, including combinations that give us a positive definite Hamiltonian. However if we wish to preserve only one $H_i$, i.e. set $c_1$ or $c_2$ to zero, then the positive definiteness is lost. The same is true for $\mathbf{Tb_2^{\pm}}$. The only difference is that we can no longer set $c_2$ to zero. Therefore, even with the well behaved non singular transformations, we can see that the choice of embedding effects whether or not we are able to recover only the original Hamiltonian $H_1$, with only $f_a$ allowing for this. 

Based on this, we can conclude that the models which give a positive definite formulation of the PU Hamiltonian have some validity, but only if considered in relation to the models Bi-Hamiltonian structure. However, as we will now discuss, the Bi-Hamiltonian structure no longer holds in the presence of interactions, at least for this model. 

\section{Interactions}
\label{interactions}

\begin{figure}
\centering
\includegraphics[scale=0.5]{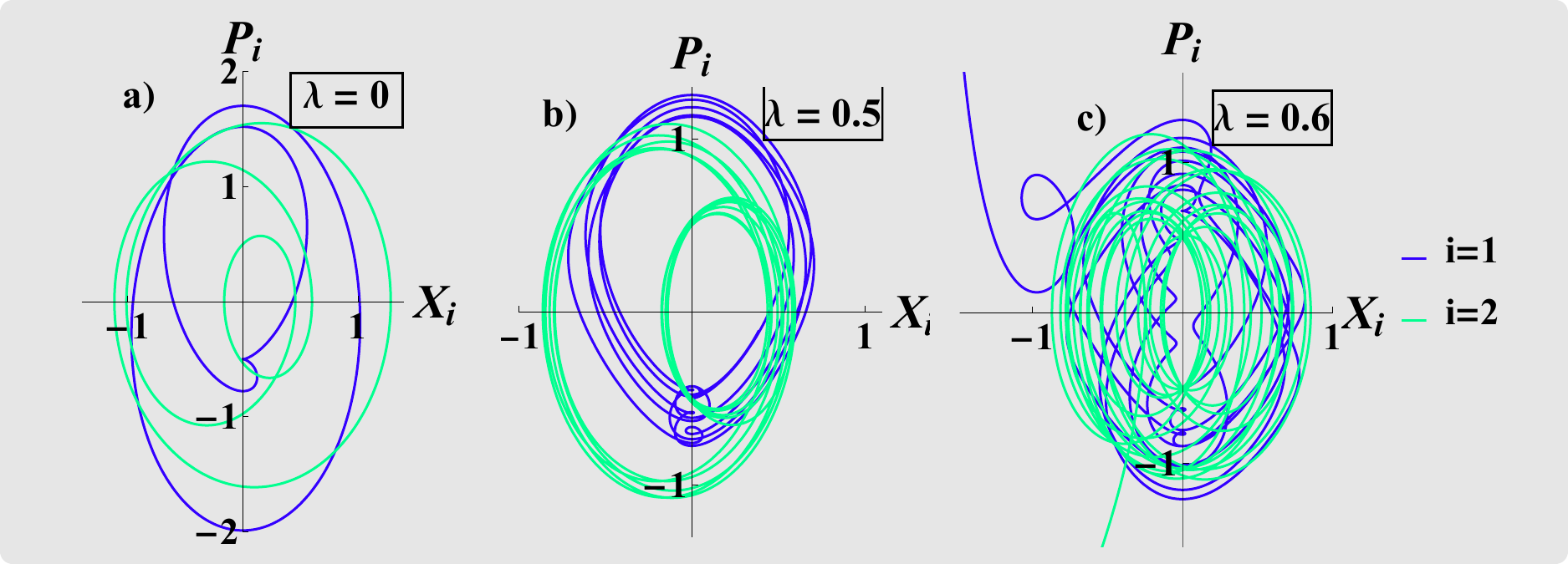}
\caption{Phase space plots for Pais-Uhlenbeck oscillator with quartic potential $W\left(q \right) = \frac{1}{4} \lambda q^4$. The coordinate $x_i$ and $p_i$ are the coordinates given in (\ref{PUphase}).  The frequencies are $\omega_1 =1$, $\omega_2=2$. The intial conditions are $x_1\left(0\right)=x_2\left(0\right)=0$, $p_1\left(0\right)=-p_2\left( 0\right)=0.5$.}
\label{figure}
\end{figure}

It is often argued that the problem of HTDTs is a problem of interactions. This is not entirely true, at the classical level some are sick even in their linear form, see the model disscussed in \cite{diezrifastaudt}. However it is true that stable HTDTs can become unstable in the presence of interactions, a good overview of this for the Pais-Uhlenbeck oscillator is given by\cite{PAV_I__2013}. The way that this is usually understood is that if a theory allows for positive and negative energies, allowing for self interaction means energy to be exchanged indefinetly between the two modes, allowing for runaway dynamics. However, as we have seen above, when it comes to the free PUO it is perfectly possible to write down a positive definite formulation, based on its Bi-Hamiltonian structure. These considertions naturally raise the question of what happens to this Bi-Hamiltonian strucutre if an interacting theory is considered. In order to see this, we consider the PUO vector field with some generic interaction term, \\
\begin{subequations}
\begin{minipage}{0.5\linewidth}
\begin{align}
L_{int} & = L -  W \left( q \right).
\end{align}
\end{minipage}
\begin{minipage}{0.5\linewidth}
\begin{align}
V_{int} & = V +  W'(q) \partial_{q^{(3)}}.
\end{align}
\end{minipage}
\end{subequations}
The question we need to answer is whether or not this vector field is locally Hamiltonian. Therefore we need to know if there is some anti-symmetric contravariant tensor, that vanishes when the Lie derivative is taken with respect to $V_{int}$. In other words, we need to solve the equation,
\begin{align}
V_{int}^{k} \frac{\partial J_{\text{int}}^{i,j}}{q^{(k)}} - \frac{\partial V_{int}^i }{q^{(k)}} J_{\text{int}}^{k,j} - J_{\text{int}}^{i,k} \frac{\partial V_{int}^j}{\partial q^{(k)}} = 0 
\label{InteractingLieDerivative}
\end{align}
note that here we are following the procedure given in \cite{Damaskinsky_2006}, and have simply added an interaction term. We find, that in contrast to the Bi-Hamiltonian strucutre found above, the only Poisson tensor satisfying equation (\ref{InteractingLieDerivative}) is given by
\begin{align}
J_{\text{int}}  & = J_1. 
\end{align}
Therefore the Bi-Hamiltonian structure of the theory is broken, and the possibility of a positive definite formualtion of this theory no longer exists. The consequences of this can be observed in the unbounded motion in the phase space plots in figure \ref{figure}. Note that this does not happen until value of coupling constant reaches a certain threshhold. It is on this basis that some authors, see e.g., make the disntinction between benign and malicous ghosts mentioned above \cite{Smilga_2005}. In this view, the ghosts of this theory are benign in the first two plots of figure \ref{figure}. The unbounded motion in the rightmost panel indicates the presence of malicous ghosts.  

\section{Conclusion}

A systematic study of the possible different representations of the Pais-Uhlenbeck oscillator shows that positive definite formulations of the free theory can be made sense of by considering the Bi-Hamiltonian structure of the model. This Bi-Hamiltonian structure can be generated by considering the Lie symmetries of the model, suggesting that this gives a systematic way to generalise this to higher $N$ \cite{felski20252}. This Bi-Hamiltonian structure of this model is, however, broken in the presence of interactions. This naturally raises the question of whether or not we can find non-linear Bi-Hamiltonian systems that can possibly be treated along similar lines as to what has been outlined here, potentially giving us a new appraoch to mitigate the ghost problem in some models. 

\section*{Acknowledgements} The work presented in this article was done in collaboration with Andreas Fring and Alexander Felski. BT would like to thank the organisers of ISQS-29 for the opportunity to present at the conference. BT is supported by a City, University of London Research Fellowship. 

\printbibliography

\end{document}